\documentclass{ws-ijmpa}
\usepackage[super,compress]{cite}
\usepackage{graphicx}
\usepackage{amssymb}
\usepackage{lineno}
\usepackage{graphicx}
\usepackage{amssymb}
\usepackage{xcolor}
\usepackage{subfig}
\usepackage{hyperref}
\usepackage{filecontents}
\usepackage{txfonts}
\usepackage{booktabs}

\newcommand{\Ec}{E_{\mbox{\scriptsize{F}}}}
\newcommand{\Eca}{E_{\mbox{\scriptsize{F}}}/(1+z')}
\newcommand{\Eb}{E_{\scriptsize{B}}}
\newcommand{\Ep}{E_{\scriptsize{Pl}}}

\begin{document}
\markboth{Kováčik, Ďuríšková, Rusnák}{Phenomenology of the dispersion law in a three-dimensional quantum space}

%
\catchline{}{}{}{}{}
%

\title{Phenomenology of the dispersion law in a three-dimensional quantum space}

\author{Samuel Kováčik}

\author{Michaela Ďuríšková \footnote{Masaryk University.}}

\author{Patrik Rusnák \footnote{Comenius University in Bratislava.}}

\address{Faculty of Mathematics, Physics and Informatics, Comenius University in Bratislava, Mlynská dolina F1\\
Bratislava, 842 48, Slovakia \\
samuel.kovacik@fmph.uniba.sk}

\address{Department of Theoretical Physics and Astrophysics, Faculty of Science, Masaryk University, Kotlářská 267/2\\
Brno, 611 37, Czechia}

\maketitle
\flushbottom

\begin{abstract}
The concept of a quantum structure underlying space has been proposed for quite some time, yet it has largely eluded direct observation. There is a prevailing notion that the minuscule effects stemming from quantum space could, under certain astronomical conditions, accumulate to an observable magnitude. Typically, these effects are perturbative, limited to the first one or two orders. This study presents a full dispersion law derived from a three-dimensional quantum space model. Our primary focus is investigating the in-vacuo dispersion phenomenon and the threshold anomaly. Furthermore, we briefly explore the implications of additional spatial dimensions within this framework and suggest a way of distinguishing between various theories that agree in the leading order approximation.

\end{abstract}

\section{Introduction}

The idea of quantum space followed shortly after the formulation of quantum mechanics; Snyder described the first mathematical model in 1947 \cite{Snyder:1946qz}. From the Planck units, we can estimate that the quantum structure of space will manifest at the scale $\lambda \approx 10^{-35}\mbox{m}$, which is many orders away from the current reach of particle accelerators or laboratory experiments. The generality of the minimal length scale feature has been reviewed by Hossenfelder \cite{Hossenfelder:2012jw}, and possible phenomenological scenarios were gathered by Amelino-Camelia \cite{Amelino-Camelia:2008aez} and a growing community recently \cite{Addazi:2021xuf}. 

In brief, a quantum structure of space seems to be a rather robust prediction and a common feature across different models. Still, it leads to effects that are exceedingly difficult to observe. There are, importantly, a few examples that seem to be within the reach of current technologies. Probably the aspect most worthy of attention is the wavelength dependant vacuum velocity that can affect measurably particles moving at the speed of light \cite{Amelino-Camelia:1997ieq, Amelino-Camelia:2008aez, Amelino-Camelia:2016ohi, Amelino-Camelia:2000cpa, Amelino-Camelia:2023srg} or close to it \cite{Amelino-Camelia:2016ohi, Amelino-Camelia:2022pja, Amelino-Camelia:2016fuh}. Recently, observation of TeV photons from the GRB 221009A brought attention to the threshold anomaly \cite{Burns:2023oxn, Li:2022wxc, LHAASO:2023lkv}; the quantum structure of space, or some other cause, reduces the interaction between gamma and low-energy background photons. This would explain the large number of detected TeV photons since their flux should be within the standard model attenuated \cite{Li:2022vgq}. Also, perhaps even more importantly, there are other possible explanations for the observed discrepancy; for example, axion-like particles are a possible explanation of the threshold anomaly \cite{Marsh:2015xka, Meyer:2013pny, Sanchez-Conde:2009exi}. In summary, we are entering an era in which astronomical observations might drive the search for physics (well) beyond the standard model. The discussed effects are present in different perspectives to that discussed in our paper, for example, the doubly-special relativity \cite{Amelino-Camelia:2002cqb}, Lorenz invariance violation \cite{Mattingly:2005re}, generalized uncertainty principle \cite{Maggiore:1993rv}, quantum loop gravity \cite{Levy:2024sdr} and others. 

Since we expect effects to be considerable for large energies, keeping only the leading corrections in $E/\Ec$ is often justified. Here, $\Ec$ is the fundamental energy scale, which might be the Planck energy scale or can be of a different order. This work presents a full dispersion law obtained from studying three-dimensional quantum space. A series of studies showed that the rotationally invariant model can lead to exact results instead of perturbative results obtained in other descriptions \cite{Galikova:2013zca, Presnajder:2014nia, Galikova:2013mia}. Also, this model was analyzed in the context of microscopic black holes \cite{Kovacik:2017tlg, Kovacik:2022ngv, Kovacik:2021qms}, and recently, its matrix formulation has been proposed, which allows for a wider range of astrophysical studies \cite{Kovacik:2023zab}. Therefore, we discuss this model in the context of quantum-space phenomenology. 

Here is the outline of the paper. Section 2 introduces the model of the mentioned three-dimensional quantum space. In Section 3, we summarize some aspects of nontrivial vacuum dispersion. In Section 4, we discuss the robustness of the predictions, and in Section 5, we conclude our findings. 

\section{Vacuum dispersion from an exact formulation of three-dimensional quantum space}

It is often assumed that space has a structure on a microscopic scale. Planck units give this scale a natural value, but in principle, different options should be considered unless we have data to pick the correct one. It can be argued, see \cite{Hossenfelder:2012jw}, that we can expect that on this scale, one cannot measure the position of a particle with arbitrary precision. This means that in quantum formalism, we have two position operators that do not commute:
\begin{equation} \label{NC}
[\hat{x}_i, \hat{x}_j] = i \hat{\theta}_{ij}.
\end{equation}
Both sides of this equation have the dimension of length squared. A reasonable choice for the r.h.s. of this equation that leads to a rotationally invariant model is given by
\begin{equation} \label{NC2}
\hat{\theta}_{ij} = 2 \lambda \varepsilon_{ijk} \hat{x}_k,
\end{equation}
where $\varepsilon$ is the Levi-Civita tensor, and $\lambda$ is a constant with the dimension of length that dictates the scale of space quantumness. An obvious feature of this relation is that in the limit $\lambda \rightarrow 0$, the ordinary quantum mechanics relation is restored, and the model describes the ordinary three-dimensional space. Therefore, $\lambda$ can be considered a deformation parameter, usually related to the Planck length. The rotational symmetry of this relation leads to the exact formula for the Coulomb problem spectrum, as has been shown in a previous work \cite{Galikova:2013zca}. 

Models that satisfy the relation \ref{NC2} can be constructed using the noncommutative star product \cite{Gracia-Bondia:2001ynb}, auxiliary bosonic operators \cite{Galikova:2013zca, Galikova:2013mia} or matrices \cite{Kovacik:2023zab}. Let us briefly discuss one way to construct a quantum space that provides insight and intuition. To quantize a theory of something, one needs to replace Poisson brackets with a commutator, for example, $\{x,p\} \rightarrow \frac{1}{i \hbar}[\hat{x},\hat{p}]$. This is usually done on a phase space, which is by default even-dimensional; however, the space $\textbf{R}^3$ is odd-dimensional. There is a workaround --- one can first embed $\textbf{R}^3$ into a higher dimensional space with a natural Poisson structure $\{z_{\scriptsize{\alpha}},\bar{z}_{\scriptsize{\beta}}\}$, quantize it there and return to $\textbf{R}^3$. As shown in \cite{Kovacik:2018vvx}, one can formulate ordinary $\textbf{R}^3$ quantum mechanics in a space with two complex dimensions, $\textbf{C}^2$, quantize that and then consider only states independent of the extra dimension. As a result, one is left with a three-dimensional theory of quantum mechanics in a quantized space. It can be shown that this theory reproduces the ordinary quantum mechanics in the $\lambda \rightarrow 0$ limit, justifying the construction. The constructed model lacks translational symmetry, so the form of the momentum conservation law cannot be established directly. 

This yields the starting kit of quantum mechanics: the Hilbert space of states $\Psi$, the position operators $\hat{x}_i$ and the Hamiltonian $\hat{H}$ obtained from the Poisson structure. From this, the velocity was defined and studied in \cite{ncqmVelo}, where the following relation between the free Hamiltonian and the velocity operator, $\hat{v}_i$, was found:
\begin{equation}
 \hat{H}_0 = \frac{ 1-\sqrt{1-m^2 \lambda^2 \hat{v}^2}}{m \lambda^2} = \frac{m \hat{v}^2}{2} + \frac{m^3 \hat{v}^4 }{8} \lambda^2+ {\cal{O}}(\lambda^4).
\end{equation}
The usual relation is recovered in the $\lambda \rightarrow 0$ limit. We can also utilize the Legendre transformation, $\hat{p}=\frac{\partial \hat{H}_{0}}{\partial \hat{v}} = \frac{\hat{m v}}{\sqrt{1-m^2 \hat{v}^{2}\lambda^{2}}}$, to find:
\begin{equation} \label{dispersion}
 \hat{H}_{0} = \frac{1}{m \lambda^{2}}\left( 1-\sqrt{1-\frac{\lambda^{2}\hat{p}^{2}}{1+\lambda^{2}\hat{p}^{2}}}\right) = \frac{\hat{p}^2}{2m} - \frac{3 \hat{p}^4}{8m} \lambda^2 + {\cal{O}}(\lambda^4).
\end{equation}
This expression was derived within nonrelativistic quantum mechanics using units for which $\hbar=c=1$; additional effects might be present in the full theory. However, it is plausible that the dispersion law sensitive to the scale of $\Ec$ will be of a similar form, and one can replace $p^2/2m \rightarrow pc$, arguing the universality of the modification as in \cite{Amelino-Camelia:2008aez}. The maximal achievable energy is $\Ec = 1/m \lambda^2$, and we will fix this to be the energy cut-off scale by taking $\lambda \rightarrow 1 / \sqrt{m \Ec}$. We will also consider only energy (momentum) eigenstates sa-tisfying $\hat{H}_0 \Psi = E \psi$ (or $\hat{p} \Psi = p \Psi$) as it was shown in \cite{ncqmVelo} that these operators commute and, therefore, can be diagonalized simultaneously. By doing these modifications and putting the expression in the common form, we obtain the square of the energy:
\begin{equation} \label{our dispersion}
 E^2(p) = \Ec^2 \left( 1 - \sqrt{1-\frac{2 p c}{ \Ec+2 pc }}\right)^2 = p^2 c^2 \left(1 - \frac{3 p c}{\Ec} + {\cal{O}}\left(\Ec^{-2}\right)\right).
\end{equation}

This form agrees, up to a numerical factor in the leading correction term, which is ambiguous due to the unfixed value of $\Ec$, with dispersions laws commonly used in previous works, for example, in \cite{Addazi:2021xuf, Li:2022vgq, Jacob:2008bw}. The leading correction is of the order of ${\cal O} \left(E/\Ec\right)$, which was expected. We take this relation to hold for any relativistic particle, however this is without the knowledge of the full theory of quantum space so changes are still possible. The particle accelerator LHC is currently probing the energies of the order of $10\ \mbox{TeV}$, which is 15 orders below the Planck energy scale --- we are not hoping to see signals of quantum space there; that is, unless the existence of extra spatial dimensions changes the value of Planck scale considerably as we discuss later. Or, as is also possible, the effective scale of space structure is some orders away from the Planck scale. However, the minuscule effect can build up for particles that travel across the universe for billions of years and can lead to measurable effects. Mixing UV and IR physics can apply a lever that enhances these phenomena. We will now discuss both of these effects in more detail. 

\section{Phenomenological aspects}
\subsection{Flight time delay} 
One of the obvious consequences of the nontrivial dispersion law, such as \ref{dispersion}, is a wavenumber-dependent wave propagation velocity. Depending on the sign in front of the first correction term, waves with higher frequencies travel either faster or slower than the low-frequency ones --- and this happens even in an empty space. Gamma-ray bursts, therefore, represent an efficient tool for studying the in-vacuo dispersion, one of the potential observable consequences of the quantum theory of gravity. Some of its (effective) models predict that quantum space-time behaves like a dispersive medium. One can envision such a situation as two photons of different energies being emitted from the same source simultaneously but detected at different times. The amplitude of the time delay is usually negligible, but the effect accumulates along cosmological distances.
If we take the dispersion law for massless particles to be of the form as assumed in \cite{Amelino-Camelia:2008aez}:
\begin{equation} \label{dispersion simple}
 p^2 \simeq E^2 /c^2 + \eta p^2 \left( E/\Ec \right)^n,
\end{equation}
the arrival-time difference relation between particles of various energies is, according to \cite{Amelino-Camelia:2008aez}, given as:
\begin{equation} \label{delta_t}
 \Delta t \simeq \eta \frac{n+1}{2H_{0}} \frac{p^{n}}{\Ec^{n}} \int_{0}^{z} dz' \frac{(1+z')^{n}}{\sqrt{\Omega_{m}(1+z')^{3} + \Omega_{\Lambda}}}.
\end{equation}
Here $\Omega_{m}$, $\Omega_{\Lambda}$, $H_{0}$ are cosmological parameters and $\Ec$ is usually taken as the energy cut-off scale --- perhaps the Planck energy --- or the scale of Lorenz invariance violation, $\eta$ is a numerical factor. Two options usually appear in the literature, $n=1,2$, where $n=1$ leads to stronger effects. While working with redshifts $z\ll1$, one can simplify the relation (\ref{delta_t}) into the following form \cite{PhysRevD.80.084017}:
\begin{equation} \label{delta_t_small_z}
 \Delta t \vert_{\mathrm{small-z}} \simeq \eta \left(\Delta E/\Ec \right)^nd/c,
\end{equation}
where $d = \frac{c z}{H_{0}}$ is the distance of the source and $\Delta E$ is the energy difference between the two considered wavelengths. 

We can now follow the strategy of \cite{Jacob:2008bw} to derive a similar formula for our dispersion law \ref{our dispersion}. We replace the momentum with the comoving momentum $p/a$, transform back to velocities by $v=\frac{\partial H}{\partial p}$:
\begin{equation}
 v(E) = c a^{-1} \left(\frac{a \Ec}{a \Ec + 2 pc}\right)^{3/2}.
\end{equation}
We can now replace the scale factor $a=\frac{1}{1+z}$ and integrate over to find that a received signal travelled:
\begin{equation}
 x(z,E) = \frac{c}{H_0} \int \limits_0^z \left(\frac{ \Eca}{\Eca + 2 pc}\right)^{3/2} \frac{dz'}{\sqrt{\Omega_m\left(1+z'^3\right)+\Omega_\Lambda}}.
\end{equation}
Importantly, if we have two photons with energies $E_1$ and $E_2$, the time difference of their arrivals will be (neglecting subleading effects):
\begin{eqnarray}\nonumber \label{delta T}
\Delta t (E_1, E_2,z) &=& \frac{c}{H_0} \int \limits_0^z \left(\left(\frac{ \Eca}{ \Eca + 2 p_1c}\right)^{3/2} - \left(\frac{ \Eca}{ \Eca + 2 p_2c}\right)^{3/2}\right) \\ 
 && \times \frac{dz'}{\sqrt{\Omega_m\left(1+z'\right)^3+\Omega_\Lambda}} \\ \nonumber
 &=& - \frac{3 c }{H_0 } \frac{\Delta E}{\Ec} \int \limits_0^z \frac{\left(1+z'\right) dz'}{\sqrt{\Omega_m\left(1+z'\right)^3+\Omega_\Lambda}} + {\cal{O}}((E/\Ec)^2),
\end{eqnarray}
where $E=pc$. This result is consistent, up to a small numerical factor, with values appearing in the literature. The integral can be expanded in $E$, and each term can be expressed using hypergeometric functions.

\subsection{Threshold anomaly} 
The important aspect of the modified dispersion law is that it affects particles propagating in the vacuum of intergalactic space. However, this medium is a vacuum only within the first approximation since it is filled, among other things, with background photons. A reversed process to electron-positron annihilation can happen as long as the energy of the photon with larger energy satisfies, assuming the trivial dispersion law, the following: 
\begin{equation} \label{trashold}
 E \geq E_{\mathrm{th}} = \frac{m_{e}^{2}}{E_{b}} , 
\end{equation}
here $m_e$ is the electron mass, $E_b$ is the energy of the background photon and $E_{\mathrm{th}}$ is the threshold energy. This follows from the conservation of four-momentum as was described, for example, in \cite{Li:2021cdz}; however, we will show a slightly different description. In \cite{Franceschini:2017iwq}, one can find reasonable estimates and distributions for energies of background photons of extragalactic origin, $E_{b}$; the other option is to consider CMB photons with $E_{b} \sim 10^{-3} \mbox{eV}$. In what follows, we will consider various values. This interaction causes attenuation of energetic photons travelling on $\mbox{Mpc}$ and $\mbox{Gpc}$ scales. Photons with energies on $\mbox{TeV}$ scales and from distant sources have been observed, which left a paradox to be explained; a possible solution is the discussed correction to the dispersion law \cite{Amelino-Camelia:2008aez, Kifune:1999ex, Aloisio:2000cm, Amelino-Camelia:2000bxx, Protheroe:2000hp}. A recent example is the observation of GRB 221009A, for which the distance was estimated to be 724 Mpc \cite{Burns:2023oxn} with peak energies reported at least at 18 $\mbox{TeV}$ \cite{Sahu:2022gvx}.

One can think of this process in the following way. The interaction of two photons with energies $E$ and $\Eb$ can create a particle-antiparticle pair with particle mass $m = \sqrt{E \Eb}/c^2$. As long as $m$ is smaller than the mass of any particle allowed by conservation laws, this interaction does not occur. Here, $\Eb$ is considered fixed and taken to be the energy of the background photons. By increasing $E$, the electron mass is reached, and pair production becomes possible; therefore, photons of sufficiently large energies disappear gradually from the flux. With the considered form of the dispersion law, \ref{our dispersion}, and other choices with a negative leading correction, the obtainable mass is, due to the finiteness of $\Ec$, reduced. That means that for sufficiently large $E$ and $1/\Eb$ is the production suppressed and the intergalactic space remains transparent, contrary to the classical prediction.

The minimal mass of a particle produced by a high-energy photon with the momentum $p$ is:
\begin{eqnarray} \label{minmass}
 m &=& \frac{1}{2c^2} \sqrt{E^2(p) -\left(pc\right)^2 + 2 \Eb \left( E(p) + pc \right)}) \\ \nonumber
 &=& \frac{1}{2c^2} \sqrt{\Ec^2 \left( 1 - \sqrt{1-\frac{2 pc}{ \Ec+2 pc }}\right)^2 -p^2c^2 + 2 \Eb \left( \Ec \left( 1 - \sqrt{1-\frac{2 p c}{ \Ec+2 pc }}\right) + pc \right)} \\ \nonumber
 &\approx & m_0 \left(1 - \frac{3 \left(pc\right)^2}{8 \Ec \Eb}\right) + ... 
\end{eqnarray}
where $m_0 = \sqrt{pc \Eb}/c^2$ would be the unmodified threshold of the pair production result. We can observe that modifying the dispersion law indeed reduces the obtainable mass and suppresses the production. The result is sensitive to the chosen value of $\Eb$. Results for some values of the background photon energy, $\Eb$ and the cut-off energy $\Ec$, are in Table \ref{table minmass}.
 
 \begin{table} 
 \centering
 \begin{tabular}{|c|ccc|}
\hline
$\Eb \setminus \Ec$ [\mbox{MeV}] & $\Ep$ & $10^{-1}\Ep$ & $10^{-2}\Ep$ \\
\hline
$1 \ \mbox{eV}$ & $4.2 $ & $3.8$ & $0.019$ \\
$10^{-1}\ \mbox{eV}$ & $1.2 $ & $0.0061 $ & $<0$ \\
$10^{-2} \ \mbox{eV}$ & $0.0019$& $<0 $ & $<0 $\\
$10^{-3}\ \mbox{eV}$ & $<0 $ & $<0 $ & $<0 $\\
\hline
 \end{tabular}
 \caption{Mass of a particle produced by an interaction between an $18 \ \mbox{TeV}$ photon with a background photon of energy $\Eb$ given the relation \ref{minmass} with various values of the fundamental energy scale $\Ec$ as measured in $\mbox{MeV} / c^2$. We can see that for theories with the fundamental scale away from the Planck length, pair production is suppressed for all background photons. For the case $\Ec \sim \Ep$, only background photons of lower energies are prevented from the interaction, as for larger values of $\Eb$, the producible mass is greater than that of electrons $m_{\scriptsize{e}} = 0.511 \ \mbox{MeV}/c^2$.}
 \label{table minmass}
 \end{table}

\begin{figure}
 	\centering \includegraphics[width=0.8\linewidth]{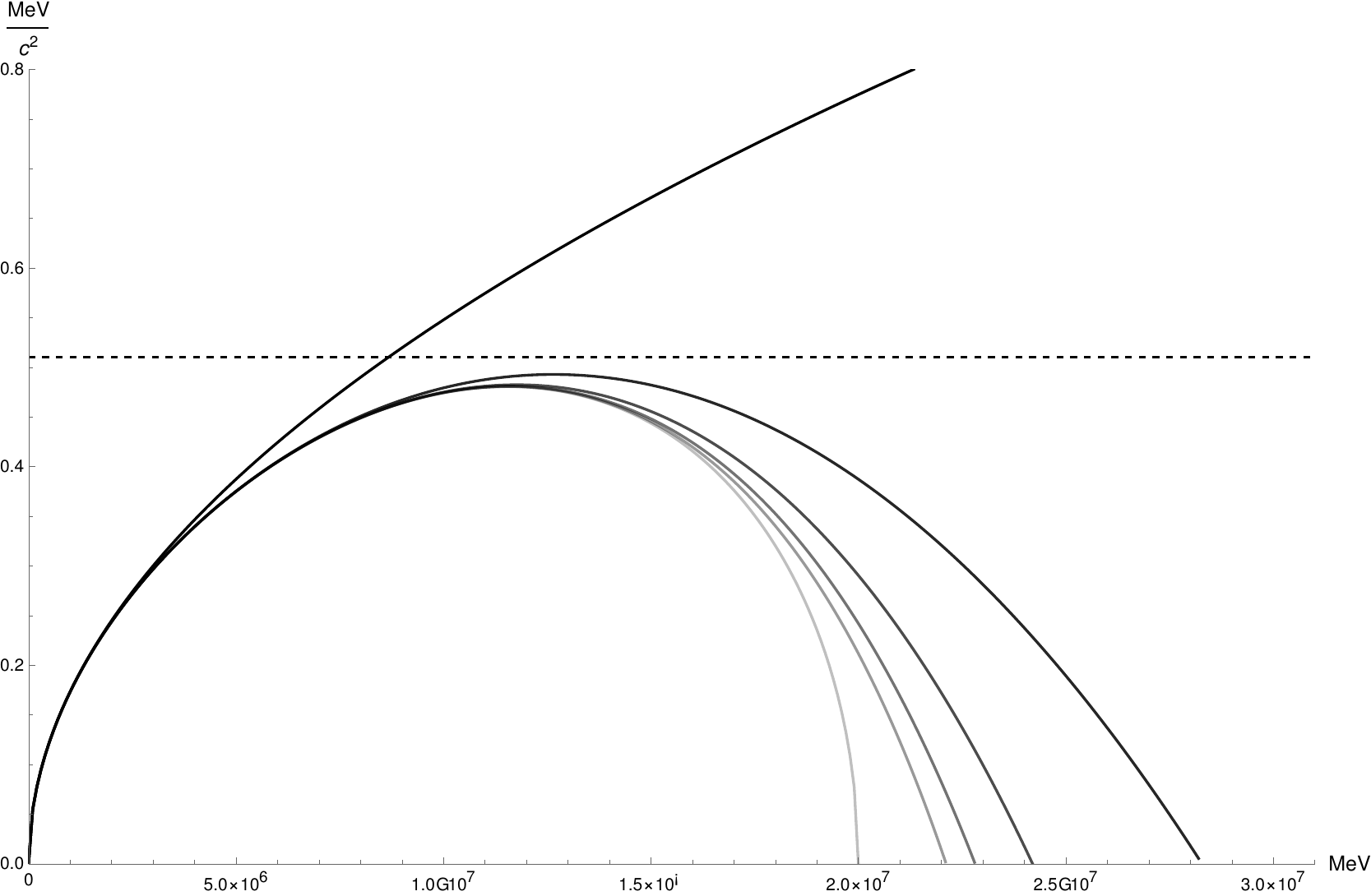} 
 	\caption{Maximal achievable mass for the classical relation from the equation \ref{minmass} with $\Ec = 10^{22} \mbox{MeV}$ and fixed $\Eb = 3 \times 10^{-2} \mbox{eV}$. Different lines correspond to different orders of expansion of \ref{minmass} in $\Ec$. The uppermost line is the classical (zeroth order) result and each line down (lighter grey) is one order higher correct; the bottom line is the exact curve; it ends at $2 \times 10^7 \mbox{MeV}$ as can be precisely calculated. The dashed line is the mass of the electron; if it is not reached, pair production is forbidden. As we can see if $\Ec \approx \Ep$, the reaction is forbidden for background photons with minuscule energies; for smaller values of $\Ec$, it is prohibitive for the larger part of the background photon spectrum, even all of it. }
 	\label{fig:ModifiedTreshold}
\end{figure}

\subsection{Constraints on extra dimensions}

The theme of our discussion so far is that there is a good reason to hope that evidence of new physics could come from astrophysical sources, even though particle accelerators have been the driving force in fundamental physics so far. A nice example is the conjectured existence of extra spatial dimensions. Nearly a decade before the launch of the Large Hadron Collider, it was suggested in \cite{Arkani-Hamed:1998jmv, Antoniadis:1998ig, Argyres:1998qn} that the existence of extra dimensions would mean that the Planck scale is shifted --- maybe even to electro-weak scale. This hypothesis has been ruled out, but it is still possible that some extra dimensions are present. The true Planck scale might still be far from the reach of particle accelerators --- but perhaps not from the observations of the effects discussed earlier in this paper. 

The basic idea is the following. While on a macroscopic scale, the ordinary Newton's gravity with force proportional to $1/r^2$ would be observed, with $D=3+d$ dimensions, one would observe $1/r^{2+d}$ dependency when measuring on a scale proportional to the extra dimension radius, $R$. The deviation from Newton's law of gravity has been searched for and not found down to the scale of approximately $50\ \mu \mbox{m}$, see \cite{Tan:2016vwu, Lee:2020zjt}. 

In space with $d$ extra dimensions of radius $R \ll r$, the gravitational potential obtains the form \cite{Arkani-Hamed:1998jmv} $V(r) \sim \frac{G_{4+d}}{r \ R^d}$. Not only does this mean that $G$ has a different value, $G_{4+d} = G_4 R^d$, --- presumable not as small as we are used to --- but also have different dimensions and therefore it relates to the Planck units differently. To be precise:

\begin{eqnarray}
  m_{\mbox{\scriptsize{pl}}, 4+d} &=& c^{\frac{1-d}{2+d}} \hbar^{\frac{1+d}{2+d}} G_{4+d}^{-\frac{1}{2+d}}, \\ \nonumber
  l_{\mbox{\scriptsize{pl}}, 4+d} &=& c^{-\frac{3}{2+d}} \hbar^{\frac{1}{2+d}} G_{4+d}^{\frac{1}{2+d}}, \\ \nonumber
  t_{\mbox{\scriptsize{pl}}, 4+d} &=& c^{-\frac{5+d}{2+d}} \hbar^{\frac{1}{2+d}} G_{4+d}^{\frac{1}{2+d}}. 
\end{eqnarray}

One can observe that if there are extra spatial dimensions, the Planck mass decreases, the Planck energy does as well, and the Planck length increases. Decreasing the Planck energy makes the vacuum dispersion effect and other effects of quantum space more prominent. This means that there cannot be that many (significantly large) extra dimensions --- as that would lead to flight-time delays, which are not being observed. This puts a constraint on the radius and number of extra dimensions, its exact value depends on precise measurements of $\Ec$.

\begin{table}
 \centering
 \begin{tabular}{|c|ccccccccc|}
 \hline
$d \setminus R[\mbox{m}]$ & $10^{-19}$ & $10^{-21}$ & $10^{-23}$ & $10^{-25}$ & $10^{-27}$ & $10^{-29}$ & $10^{-31}$ & $10^{-33}$ & $10^{-35}$ \\
\hline
$1$& $22.8$ & $23.5$ & $24.2$ & $24.8$ & $25.5$ & $26.2$ & $26.8$ & $27.5$ & $28.2$ \\
$2 $& $20.2$ & $21.2$ & $22.2$ & $23.2$ & $24.2$ & $25.2$ & $26.2$ & $27.2$ & $28.2$  \\
$3 $& $18.6$ & $19.8$ & $21.0$ & $22.2$ & $23.4$ & $24.6$ & $25.8$ & $27.0$ & $28.2$  \\
$6 $& $16.2$ & $17.7$ & $19.2$ & $20.7$ & $22.2$ & $23.7$ & $25.2$ & $26.7$ & $28.2$  \\
$9 $& $15.2$ & $16.8$ & $18.4$ & $20.1$ & $21.7$ & $23.3$ & $25.0$ & $26.6$ & $28.3$  \\
\hline
 \end{tabular}
 \caption{This table shows the value of $\log_{10} E_{\mbox{\scriptsize{pl}}, 4+d}$ with $E$ expressed in $\mbox{eV}$ for various radius sizes $R$ and different number of extra dimensions, $d$. We can see that adding some small extra spatial dimensions reduces $\Ec$ from $\Ep$ down to scales suggested by studies of Lorenz invariance violation \cite{Xiao:2009xe,Lorentz:2016aiz, Bolmont:2010np}.}
 \label{tableeps}
\end{table}

\section{Observational considerations}

The starting point of our analysis was a model of three-dimensional quantum space and a Hamiltonian operator that was the direct result of the quantisation process. Naturally, this led to a dispersion relation, which we, with some assumptions, turned into the form \ref{our dispersion}. When taking only the leading correction term, our expression agrees with some other values used in literature, \cite{Amelino-Camelia:2008aez, Huang:2018ham}. Even in the case of nontrivial vacuum dispersion, it was unclear whether to consider a leading correction linear or quadratic in $E/\Ec$; our analysis favours the former. This is important because with it, \ref{delta T} produces an observable effect. Going to the next-to-leading term leads, however, to corrections that are below the current detection threshold.

Let us compare our result with the one reported from the analysis of the recent event GRB 221109A \cite{2023JPhG...50fLT01Z}. The source of this signal was close, $z=0.1505$, but due to unprecedented energies, this observation is relatively also relevant in the context of threshold anomaly. The mentioned work operated with the dispersion relation:
\begin{eqnarray}
 v(E) = c \left(1 - \frac{n+1}{2}\left( \frac{E}{\Ec} \right) ^n \right),
\end{eqnarray}
where the case close to ours has $n=1$. The leading order correction differs by a factor of $3$, which, being ${\cal{O}}(1)$, is negligible as it comes with the undetermined energy scale $\Ec$. The value used for energy difference in this reference was $99.3\ \mbox{GeV}$, which led to the prediction of $\delta t \approx 20\ \mbox{s}$. Using the same values of $\Ec, E$ and cosmological parameters $H_0 = 67.3 \ \mbox{km}\ \mbox{s}^{-1} \mbox{Mpc}^{-1},\ \Omega_m = 0.315, \Omega_\Lambda = 0.685$ we obtain $\delta t \approx 60\ \mbox{s}$. We could obtain the same value by considering a slightly different value of $\Ec$. If we use $\Ec \approx \Ep$, the time-delay value would be $\Delta t \approx 1.83 \mbox{s}$, which is easily measurable if we can identify a low-energy photon that escaped from the source simultaneously with a high-energy one. Some efforts along those lines were made in the aforementioned reference. 

For a small satellite detector, \textit{GRBAlpha} \cite{Pal:2023qjo} for this comparison, with the upper energy limit on the order of $1\ \mbox{MeV}$, the time shift with $\Ec \approx \Ep$ would be on the order of $10^{-5} \mbox{s}$, which is below the temporal resolution of $\approx 1\ \mbox{s}$. That means that not observing the effect across multiple channels restricts the value of the fundamental energy scale, $\Ec/\Ep > 10^{-5}$. We can compare this with the values in Table \ref{tableeps} and conclude that under the aforementioned assumptions, even this small satellite disfavours extra spatial dimensions larger than $\approx 10^{-20}\ \mbox{m}$, a stronger constrain than the one obtained in LHC \cite{Deutschmann:2017bth}. Also note in Table \ref{telescopes} that some current and future telescopes are well prepared to observe this effect, especially by cross-correlation across multiple channels.

\begin{table}[h]
\centering
\begin{tabular}{|c|cc|}
 \hline
& Temporal resolution & Energy range \\ \hline
GRBAlpha\cite{Pal:2023qjo} & $1$ s & $80$ keV $-\ 950$ keV \\
Fermi GBM \cite{2022hxga.book...29T} & 2 $\mu$s & $8$ keV $-\ 40$ MeV \\
Fermi LAT \cite{2022hxga.book...29T} & $<$ $1$ $\mu$s& $20$ MeV $-\ 300$ GeV \\
Hermes \cite{2020SPIE11444E..1RF, HERMES:2021wpj} & $\leq$ 250 ns &$5$ keV $-\ $ $0.5$MeV\\ 
 \hline
\end{tabular} 
\caption{Technical limits of some GRB detectors.}
\label{telescopes}
\end{table}

The second important effect is the reduced attenuation of energetic photons due to pair production. Here, the effect is inversely proportional to the energy of background photon $\Eb$, which is tantalizing as it can make event next-to-leading terms relevant. In Figure \ref{fig:ModifiedTreshold}, we compare the minimal producible mass with the expression truncated to one and two levels ––– the difference is obvious. This means that the effect of non-trivial dispersion law is observable beyond the leading order, given one has access to the interaction of two photons of vastly different energies; this can be either relevant for astrophysical observations or maybe even in some future laboratory settings \cite{Budnev:1975poe, Esnault:2021eyg}. The laboratory experiments would be preferable because the distribution of background photon energies across cosmological scales might be challenging. Theoretically, one needs a stable source of gamma photons to traverse an isolated box filled with low-frequency radiation to make the threshold \ref{trashold} relevant.

Another interesting option is to analyse pulsar sources that might not be so distant but are stable and allow for signal accumulation. For example, it has been reported that the Vela pulsar is a source of $~ 20\ \mbox{TeV}$ photons \cite{HESS:2023sxo}. The time it takes for light to travel from this source is $~ 3 \times 10^{10}\ \mbox{s}$, and the spinning frequency of the source is $~ 10\ \mbox{Hz}$. Assuming $\Ec$ is on the Planck scale and taking $\Delta E \sim 10\ \mbox{TeV}$, this translates into flight time delays of $10^{-5} \mbox{s}$ --- which should be within resolution limits. Given the stability of the source, one could hope for a robust dataset on which a cross-correlation analysis could be done to observe an energy-dependent time delay.

\section{Conclusion}

We have touched upon several open questions in physics, with the most significant one being whether space possesses a quantum structure. If so, what structural equation does describe it? And what is the value of the parameters in it? We assumed the answer to the first question is positive and have derived a dispersion law from a robust model of three-dimensional quantum space. Some dispersion laws in the literature have a linear correction term, and some have a quadratic correction term. Our result agrees with the linear form, which leads to stronger observational effects. 

Their scale depends on the energy value at which the underlying physics becomes relevant. The natural value of this would be the Planck energy scale, but other values have been considered in the literature, somewhere between 3 and 4 orders below the Planck scale. For example, some approaches to the string theory give the value of $10^{11} \mbox{GeV}$, eight orders below the Planck scale \cite{Antoniadis:1999rm}. If the energy cut-off scale was also of this order, the flight-time delays would be not relevant only for GRB photons but also for photons of energy multiple orders lower. For larger energies, this would lead to delays on the order of years instead of seconds. Therefore, it is safe to assume we should be operating roughly in the interval $10^{-8} \Ep < \Ec \le \Ep$, perhaps somewhere in the upper half (on the logarithmic scale). 

Observing some phenomenological aspects of modified dispersion law can, in principle, be expected with reasonable confidence as various theories contain this feature. It should also be noted that some theories, for example, quantum spaces created by a star-product, have a trivial dispersion law \cite{Szabo:2001kg}. The distinction line between those remains elusive. That means that confirming the nontrivial corrections to the dispersion law would suggest some of the novel theories might be correct; not observing the deviation from the dispersion law would not disprove them all. This might be a frustrating state of matter, but on the contrary, it is astonishing that we have a window into physics well behind the reach of particle accelerators. 

The novel aspect of our research is that we have obtained a result in agreement with others used in the literature from a well-formulated and studied model of three-dimensional quantum space. The mathematical structures of this model have been studied thoroughly \cite{Galikova:2013zca, Presnajder:2014nia, Kovacik:2018vvx} and so have been some phenomenological aspects, such as the behaviour of regular black holes \cite{Kovacik:2017tlg, Kovacik:2021qms} or confined particles \cite{Bukor:2022xjc, Kovacik:2016alm}. In this regard, this paper studies new phenomenological aspects of this model. From this point of view, the results presented connect to other areas in physics and help build a foundation of an interconnected phenomenological web that is eventually needed for establishing or disproving a new theory. 

The flight-time delay we have derived agrees with previous works and depends mostly on the energy scale on which the modification to the dispersion law becomes important. Our considered correction is first-order, and the phenomenon manifests without invoking artificially low energy scales. However, in the case of threshold anomalies, we have observed different results by including various levels of approximation, which means this effect can be used to distinguish between various theories with nontrivial vacuum dispersion. Perhaps, in the future, laboratory tests of the interaction of high and low-energy photons with the required precision might be feasible. 

Regarding observations and experiments, we are in a state where brinks of new physics have been suggested by data \cite{Amelino-Camelia:2016ohi}, and planned satellite missions should solidify --- or disprove --- this evidence. Therefore, attention should be paid to all models of quantum space and other new physics features that can be verified in the near future. 

The levers of astronomical data --- either the accumulation of minuscule effects or interaction with low-energy background photons --- offer better hope of finding signals of Planck-scale physics than particle accelerators. To prove this point, we have made a little detour to verify what the known results infer about the extra spatial dimensions under similar assumptions when discussed in the context of particle accelerators. We have seen that if the presence of extra dimensions indeed caused a different value of the Planck scale, and this, in turn, would affect the scale of deviation from nontrivial dispersion law, the coherency of typical gamma signals can put severe constraints on the number and size of extra dimensions --- order of magnitude more restrictive than particle accelerators. Extra spatial dimensions in the context of GZK limit were recently discussed in \cite{Montero:2022prj}.

Astronomical searches for physics well behind the standard model are impressive. With this research, we hope to have added a new piece to the overall mosaic. 

\section*{Acknowledgement}
This research was supported by VEGA 1/0025/23 and the MUNI Award for Science and Humanities, funded by the Grant Agency of Masaryk University. We want to thank Benedek Bukor, Matej Hrmo, Juraj Tekel, Veronika Gáliková, Yuhma Asano, Peter Prešnajder, Wladimiro Leone, and Norbert Werner for their valuable comments and discussion. 

\bibliographystyle{unsrt}
\bibliography{references}

\end{document}